# Uncertainty-based information extraction in wireless sensor networks for control applications

Bartłomiej Płaczek* and Marcin Bernaś

Institute of Computer Science, University of Silesia, Będzińska 39, 41-200 Sosnowiec, Poland

**Abstract.** Design of control applications over wireless sensor networks (WSNs) is a challenging issue due to the bandwidth-limited communication medium, energy constraints and real-time data delivery requirements. This paper introduces a new information extraction method for WSN-based control applications, which reduces the number of required data transmissions to save energy and avoid data congestion. According to the proposed approach, control applications recognize when new data readings have to be collected and determine sensor nodes that have to be activated on the basis of uncertainty analysis. Processing of the selectively collected input data is based on definition of information granules that describe state of the controlled system as well as performance of particular control decisions. This method was implemented for object tracking in WSNs. The task is to control movement of a mobile sink, which has to reach a target in the shortest possible time. Extensive simulation experiments were performed to compare performance of the proposed approach against state-of-the-art methods. Results of the experiments show that the presented information extraction method allows for substantial reduction in the amount of transmitted data with no significant negative effect on tracking performance.

## 1. Introduction

Recent advances in wireless sensor networks (WSNs) have resulted in new possibilities of data collection for a wide range of control applications including home automation, environmental protection, precision agriculture, transportation systems, space exploration, industrial automation, military systems, etc. [1, 2, 3, 4]. The WSNs offer numerous advantages, such as cost effectiveness, simple deployment, mobility, high monitoring precision and vast area coverage. However, the design of control applications over WSNs raise great challenges due to the bandwidth-limited communication medium, energy constraints, data congestion and transmission delays. These issues are particularly important when dealing with control tasks that require reliable real-time data delivery [5].

Current research efforts, that address the needs of timeline and accurate data collection in WSNs, fall into two main categories. The first category includes communication protocols that are designed to maximize the data transmission performance of WSNs. The second category comprises data collection methods aimed at minimization of the demand for data transmission. Such methods optimize the use of sensor nodes by selecting data readings that are necessary for extraction of needed information. These two approaches can work in tandem to fulfill requirements of a particular application. In the literature, much work has been done to develop communication protocols for control applications of WSNs, however, relatively little research has been focused on appropriate information extraction schemes.

---

* Corresponding author. E-mail addresses: placzek.bartlomiej@gmail.com (B. Płaczek), marcin.bernas@gmail.com (M. Bernaś).

The available methods of information extraction have been devised mainly for monitoring applications [6] that require the WSN to provide information describing monitored parameters with a defined, constant precision or to report a predetermined set of events. In case of control applications, the required scope and precision of the delivered information change during the control process. They depend on past and current state of the controlled system as well as on the available control actions that can be implemented at a given time. Therefore, the dynamically changing requirements on input information have to be taken into consideration when designing the information extraction procedures for control applications based on WSNs.

The main objective of the presented study is the development of a method which allows the WSN-based control applications to recognize situations when new data have to be collected as well as to determine sensor nodes that have to be activated. The originality of the proposed method lies in the uncertainty analysis used to extract information which is necessary and sufficient for making control decisions. The uncertainty of control decision was defined as an estimate of probability associated with cases where the control decision is not optimal. According to the proposed approach, input data are delivered from WSN if it is expected that such operation will result in decrease of the decision uncertainty. Selection of the activated sensor nodes is based on determination of a data set, which is necessary to confirm or exclude the possibility that the considered control decision is not optimal. Processing of the selectively collected input data is based on definition of information granules [7] that describe state of the controlled system as well as performance of particular control decisions. The concept of information granules provides convenient representation of the extracted information, the precision of which changes during execution of control tasks.

The paper is organized as follows: Section 2 includes review of related research and describes main contribution of this paper in the context of WSN literature. In Section 3, the method of uncertainty-based information extraction is introduced in details and an illustrative example is given to demonstrate the proposed approach. Section 4 contains results of an experimental study on information extraction in WSN for mobile object tracking. The task of the considered object tracking application is to control the movement of a mobile sink which has to reach the target in the shortest possible time. In this study, the performance of the proposed method was compared against results obtained for state-of-the-art algorithms. Finally, in Section 5, conclusions are presented and some future research directions are outlined.

## 2. Related work and proposed approach

A number of approaches have been introduced in the literature that deal with information extraction in wireless sensor networks. A survey and classification of these approaches can be found in [8]. This section provides a concise description of selected information extraction methods that are relevant for control applications and focus on reducing amounts of data transmitted by sensor nodes.

### 2.1. Data aggregation

One of the fundamental techniques is based on in-network data aggregation [9]. According to this approach each sensor node at a predetermined path aggregates data transferred from other nodes in communication range, and communicates only the aggregated information to the next node in the path. Depending on method, the data aggregation path can be created randomly or can be optimized for a given query. Selectivity of queries and spatial correlations in sensor readings are taken into account by the PDT algorithm [10], which creates a data aggregation tree that minimizes the use of non-selected sensor nodes. Main disadvantage of this type of methods is a delay in data transmission, which is a consequence of the time-consuming data aggregation operations executed by intermediate nodes.

## 2.2. Data suppression

Suppression based methods make use of the fact that different observed states of the monitored physical phenomena are temporally as well as spatially correlated [11]. Temporal suppression is the most basic method: sensor readings are transmitted only from those nodes where a change occurred since the last transmission [12]. Spatial suppression reduces redundant transmissions of sensor readings from neighboring nodes that have the same or similar values [13]. In [14] a combined spatio-temporal suppression algorithm was introduced that considers the sensor readings and their differences along transmission paths to suppress reports from individual nodes.

In case of model-based suppression methods the divergence between actual sensor readings and model predictions is analyzed to detect the necessity of data transfers [15]. This approach uses a pair of models of the monitored phenomenon. The first model is used at the sink and the second one is distributed in the sensor network. Both models predict the same values of sensor readings. Sensor node sends the data only if the divergence between predicted and measured value is above a predetermined error tolerance threshold. This technique guarantees that all values available at the sink are within a fixed error bound from the measured values.

## 2.3. Model-based querying

Another approach to the problem of information extraction in sensor networks is the model-based querying approach, in which sensor readings are complemented by a probabilistic model of the monitored process [16]. Sensors nodes are used to collect data only when confidence of information provided by the model is under a required level. The confidence bounds have to be defined by the user. Data collection plan is optimized to minimize sensing and transmission cost. During data collection only those sensor readings are transmitted that are necessary to deliver information with an acceptable confidence. This approach was applied to information extraction in sensor network for light control in buildings [17].

## 2.4. Our contributions

The method proposed in this paper enables recognition of situations when information provided by a model of the controlled system is not sufficient to make optimal control decisions. According to the proposed approach, sensor data are transmitted from nodes to the sink only at selected time steps of the control procedure. For the remaining time intervals, the data transmission in WSN ceases and the sink approximates the state of the controlled system on its own using the predictive model. In such time periods the sink does not communicate with other nodes. This approach exploits the fact that control applications can tolerate approximate predictions of their input parameters. Nevertheless, uncertainty of control decisions must be appropriately low to ensure the optimal performance of a control application. Thus, the uncertainty of control decisions is taken into account to determine when the data have to be delivered from sensor nodes. The presented method evaluates uncertainty associated with the answer to question about optimality of a chosen control action. In order to minimize this uncertainty the application collects data that enable verification of such possible states of the controlled system for which the chosen control action is not optimal. The above principle is used in the selection of sensor nodes that have to communicate their readings.

The basic concept of using decision uncertainty for information extraction in WSNs has been presented in earlier publications devoted to urban traffic control [18, 19]. In those works a threshold of the decision uncertainty was used to decide when sensor data need to be collected from a vehicular sensor network. This paper extends and improves on the previous approach by introducing selection of sensor nodes and estimation of expected decrease of the decision uncertainty. It should be also noted that the method presented here is general and can be applied to a wide range of control tasks.

On the basis of the proposed method one can design information extraction algorithms that are executed on the application layer of WSN. Further optimization of the data transfers can be performed concurrently on the networking layer, e.g. by using data aggregation or data suppression techniques. It means that the proposed approach is complementary to the existing methods that were mentioned above in this section.

Comparing with the model-based querying approach, the main novelty of the proposed method lies in detecting the necessity of data transfers on the basis of uncertainty of control decisions instead of the uncertainty of measured parameters. This difference is crucial in the case of control applications, because there is no functional dependency between the uncertainty of approximation of input parameters and the uncertainty of control decision. In other words, for a given uncertainty level of parameters approximation, the uncertainty of control decision may be quite different depending on the previously extracted information and the set of available control actions that can be implemented. Therefore, it would be difficult to determine acceptable uncertainty limits for all the monitored parameters of the system under control. The proposed method overcomes this problem by analyzing expected decrease of the uncertainty of control decision.

## 3. Uncertainty-based information extraction

Control application needs input information about current state of the system under control to make decisions (to select an optimal action). The objective of the presented method is to efficiently extract the necessary information from sensor network. The method approximates minimal set of sensor nodes that provide information which is precise enough for finding the most effective control action.

### 3.1. Control decisions

We will assume that the control decisions are made in discrete time steps. At a particular time step $t$, the task of control application is to select and implement the control action $a_t \in A$, which minimizes an objective function $f$:

$$f : A \times U \to \mathbf{R}, \qquad (1)$$

where:
$A$ – set of available control actions,
$U$ – multidimensional space of parameters that describe state of the system under control,
$\mathbf{R}$ – set of real numbers.
It means that for the current state of the controlled system $x_t \in U$ the selected control action $a_t$ should satisfy the following condition:

$$f(a_t, x_t) = \min_{a \in A} \{ f(a, x_t) \}, \qquad (2)$$

In the proposed method an information granule $S_t$ is used to describe the state of the controlled system, which is recognized by the control application in time step $t$. The information granule $S_t$ is defined as a set of possible system states by using the following membership function:

$$\mu_{S_t}: U \to [0;1]. \qquad (3)$$

The above definition takes into account a situation when the available information does not identify the state of controlled system precisely. The underlying insight is that in many cases the precise information is not necessary for making control decisions. It should be also noted that this definition allows us to utilize different formal frameworks for the representation of information granules (crisp sets, fuzzy sets, rough sets, etc.) [7].

Current state of the controlled system $x_t$ has to be approximated by granule $S_t$, such that $x_t \in S_t$. The information granule $S_t$ is formed on the basis of two elements: information extracted from sensor network and prediction made by using a model $m$ of the system

$$S_t = m[S_{t-1}, I(D_t)], \qquad (4)$$

where $I(D_t)$ denotes data delivered from a set of sensor nodes $D_t \subseteq D$ and $D$ is the set of all sensors nodes in the network that can be used to measure parameters of the system.

Value of the objective function $f(a_t, x_t)$ cannot be precisely evaluated because of two reasons: firstly, the exact state of the system $x_t$ is assumed to be unknown and secondly, the evaluation is based on a prediction of future results. Therefore, we will use another information granule $F(a_t, S_t)$ to describe possible values of the objective function for a given control action $a_t$. Granule $F$ can be estimated on the basis of the granule $S_t$, by using function (1). Thus, the membership function associated with this information granule is defined as:

$$\mu_F: \mathbf{R} \to [0;1]. \qquad (5)$$

### 3.2. Uncertainty of control decision

Let us consider two alternative control actions $a'$ and $a''$. In order to decide if the control action $a'$ should be implemented at a given time step, the following condition has to be verified:

$$F(a', S_t) \leq F(a'', S_t). \qquad (6)$$

Due to using the information granules $F$, result of the above comparison does not have a Boolean value. Therefore, we will express this result in terms of probability. According to the proposed approach, the condition (6) is assumed to be satisfied if

$$P[F(a', S_t) \leq F(a'', S_t)] \geq 0.5, \qquad (7)$$

where $P[.]$ denotes probability calculated on the basis of the method presented in [20].

We employ the following formula to determine confidence level of satisfying condition (6):

$$CON[F(a', S_t) \leq F(a'', S_t)] = P[F(a', S_t) \leq F(a'', S_t)] - P[F(a', S_t) > F(a'', S_t)], \qquad (8)$$

or simplified:

$$CON[F(a', S_t) \leq F(a'', S_t)] = 1 - 2P[F(a', S_t) > F(a'', S_t)]. \qquad (9)$$

Consequently, in order to quantify the chance that the condition will not be met, we define uncertainty of the comparison result:

$$UNC[F(a', S_t) \leq F(a'', S_t)] = 1 - CON[F(a', S_t) \leq F(a'', S_t)] = 2P[F(a', S_t) > F(a'', S_t)]. \qquad (10)$$

It should be noted that the above uncertainty measure takes values from 0 to 1 if the analyzed condition is satisfied (see eq. (7)).

The procedure of forming control decision, i.e. selecting of the most effective control action $a_t$ according to eq. (2), requires a number of comparisons. Uncertainty of control decision is evaluated as a maximum of the uncertainties associated with particular comparisons:

$$UNC(a_t, S_t) = \max_{a \in A - a_t} \{UNC[F(a_t, S_t) \leq F(a, S_t)]\} \ . \tag{11}$$

### 3.3. Expected decrease of uncertainty

The underlying concept of our method is to execute the data collection in WSN only if it is expected that new data will reduce uncertainty of control decision. Therefore, we define expected decrease of uncertainty $\Delta UNC$, which is used to decide if the data collection is necessary.

At each time step of the control procedure, when current sensor readings are yet unknown, we can predict values of the decision uncertainty that will be obtained if we collect the data from all sensor nodes in the network ($D$). To this end, all possible sets of sensor readings $\hat{I}(D)$ should be taken into account. According to eq. (4) each data set $\hat{I}(D)$ forms an information granule:

$$\hat{S}_t = m[S_{t-1}, \hat{I}(D)], \tag{12}$$

which describes a hypothesis about current state of the controlled system. In this way a family of information granules $\hat{\mathbf{S}}_t$ is obtained. It should be noted here that the family of information granules $\hat{\mathbf{S}}_t$ corresponds to the most precise approximation of the system state, which can be achieved by collecting all the data that are available in sensor network. It is also important that different control decisions $\hat{a}_t$ may be made depending upon the information granule $\hat{S}_t$ under consideration.

For every granule in family $\hat{\mathbf{S}}_t$ the uncertainty of control decision $UNC(\hat{a}_t, \hat{S}_t)$ can be predicted by using eq. (11). Let us denote the average value of the predicted uncertainties by $\overline{UNC}(\hat{\mathbf{S}}_t)$. This measure allows us to evaluate expected decrease in the decision uncertainty, which will be observed if the input data are delivered from sensor network:

$$\Delta UNC(a_t, S_t^\varnothing) = UNC(a_t, S_t^\varnothing) - \overline{UNC}(\hat{\mathbf{S}}_t) , \tag{13}$$

where

$$S_t^\varnothing = m[S_{t-1}, \varnothing] \tag{14}$$

represents the least precise approximation of current state of the controlled system, obtained without data from sensor network.

According to the introduced method, the definition (13) is used to decide if the data collection should be executed at a given time step of the control procedure. The input data are collected only if the expected decrease of uncertainty $\Delta UNC(a_t, S_t^\varnothing)$ is above a predetermined threshold $\alpha \in [0;1]$.

### 3.4. Selection of sensor nodes

After deciding that the data collection is necessary, the next problem to be solved is the selection of sensor nodes that will be requested to send their data readings. In this paper two different strategies of the sensor nodes selection will be examined: the *exhaustive strategy*, which aims in verification of all possible states of the controlled system, and the *minimum strategy*, which verifies only those states where the analyzed control decision would be non-optimal.

In the following part of this section we will use index *j* to uniquely denote an information granule. Each information granule $\hat{S}_t^{(j)} \in \hat{\mathbf{S}}_t$ represents a hypothetic state of the controlled system, which is described by data set $\hat{I}(D)^{(j)}$ that consist of possible values $y_{i,t}^{(j)}$ of the parameters measured by sensor nodes $i \in D$:

$$\hat{I}(D)^{(j)} = (y_{1,t}^{(j)}, \ldots y_{n,t}^{(j)}), \quad (15)$$

where $D = \{1, \ldots n\}$.

According to the exhaustive strategy, we need to take into account such parameters that uniquely describe all the hypothetic states of the controlled system. In other words, the objective is to find the minimal set of sensor nodes $D_t \subseteq D$, for which the following condition will be satisfied:

$$\forall (\hat{S}_t^{(j)}, \hat{S}_t^{(k)}) \in \hat{\mathbf{S}}_t \; \exists i \in D_t : y_{i,t}^{(j)} \neq y_{i,t}^{(k)}. \quad (16)$$

It means that the sets of sensor readings forming granules $\hat{S}_t^{(j)}$ and $\hat{S}_t^{(k)}$ have to include at least one parameter which will take different values distinguishing them.

Using the minimum strategy, we require the selected sensor nodes $D_t$ to provide data that will enable verification of all the hypothesis represented by information granules $\hat{S}_t$ for which the control decision $\hat{a}_t$ (selected control action) is different than the control decision $a_t$ made on the basis of information granule $S_t^\varnothing$. Let us denote by $\hat{a}_t^{(j)}$ the control action selected as an optimal one on the basis of the information granule $\hat{S}_t^{(j)}$. We are interested in finding the minimal set of sensor nodes $D_t \subseteq D$, for which the following condition is fulfilled:

$$\forall (\hat{S}_t^{(j)}, \hat{S}_t^{(k)}) \in \hat{\mathbf{S}}_t : \hat{a}_t^{(j)} \neq \hat{a}_t^{(k)} \; \exists i \in D_t : y_{i,t}^{(j)} \neq y_{i,t}^{(k)}. \quad (17)$$

It means that for each pair of information granules $(\hat{S}_t^{(j)}, \hat{S}_t^{(k)})$, such that $\hat{a}_t^{(j)} \neq \hat{a}_t^{(k)}$, the corresponding states of the controlled system can be distinguished from one another based on the parameters available in $\hat{I}(D_t)$.

In practice, implementation of the above strategies requires algorithms designed and optimized for a specific application. One general remark is that the problem of sensor nodes selection can be solved by using the algorithms for reducts evaluation, that are available in the theory of rough sets [21].

### 3.5. An illustrative example

An example of a predator-prey problem will be used to illustrate the above formulation of the information extraction method. For the sake of clarity, the example is restricted to a single decision at time step $t$. Let us assume that the objective of the control is to minimize the time in which the predator catches a prey. We will consider a case in which two preys are in range of the predator. The predator has to decide which prey to go for. Thus, the possible control decisions are: to chase the first prey ($a_t = 1$) or to chase the second prey ($a_t = 2$). State of the controlled system $x_t$ is described by distances between the predator and the two prey: $x_t = (x_t^{(1)}, x_t^{(2)})$, where $x_t^{(1)}$ denotes distance between predator and the first prey, and $x_t^{(2)}$ is the distance between predator and the second prey at time step $t$. The distances can be determined on the basis of data delivered from sensor nodes in WSN. Each sensor node detects prey in a segment of 10 m. Available information, which is obtained at time step $t$ from a model of the prey motion, says that the possible distances to the first and second prey are in intervals (10,50) [m] and (30,70) [m] respectively. This information is represented by the following granule (hyperbox):

$$S_t^\varnothing = (10,50) \times (30,70). \quad (18)$$

The problem under consideration is illustrated in Fig. 1. Let us assume that velocity of the predator equals 30 m/s and the possible velocities of the prey are contained in interval (20,25) [m/s]. Thus, the time to catch can be predicted as: $F(1, S_t^\varnothing) = (1,10)$ [s] for the first control action and $F(2, S_t^\varnothing) = (3,14)$ [s] for the second control action (i.e., for decision to chase the second prey).

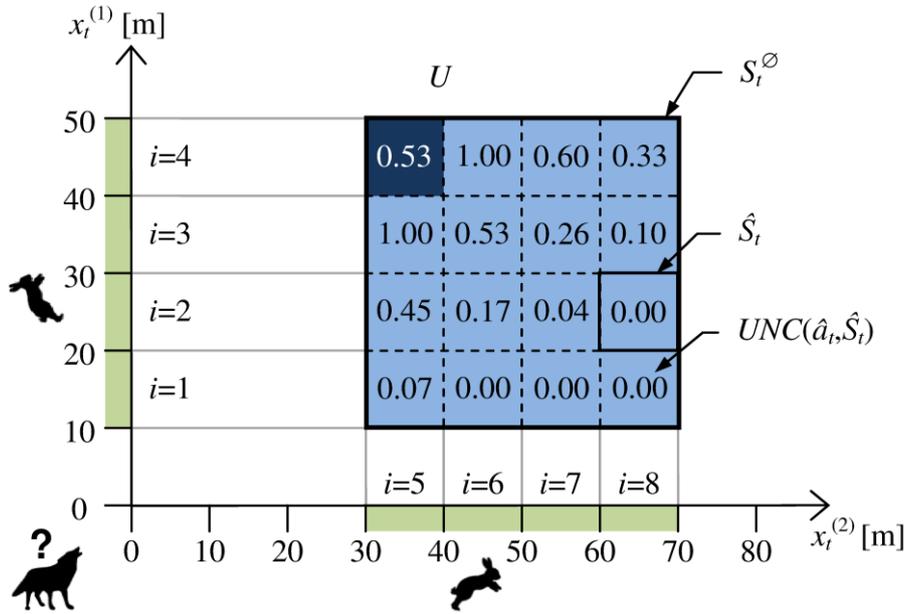

Fig. 1. Information granules and uncertainty of predator decision

Using the probabilistic approach to intervals comparison [20] we can compute the probability $P[F(1, S_t^\varnothing) \leq F(2, S_t^\varnothing)] = 0.755$. Since the probability is above 0.5, the first control action becomes the candidate for implementation ($a_t = 1$). In order to check the necessity of data delivery from WSN, we calculate the decision uncertainty $UNC(1, S_t^\varnothing) = 0.49$ as well as the predicted average uncertainty $\overline{UNC(\hat{S}_t)} = 0.32$. On this basis the expected decrease of uncertainty is determined: $\Delta UNC(1, S_t^\varnothing) = 0.17$. Details of the calculations are presented in Tab. 1.

The chart in Fig. 1 shows information granules (hyperboxes) that represent hypothetical states of the controlled system. The numbers within squares indicate decision uncertainties related to the particular information granules. Each information granule has a corresponding tuple of sensor readings $\hat{I}(D)$, presented in Tab. 1. We take into account only the eight sensor nodes that cover the area where prey can be detected. Value 1 of a sensor reading denotes that presence of the prey is detected by a given sensor node.

In the analyzed example, the predetermined threshold value for the expected decrease of uncertainty ($\alpha = 0.1$) is exceeded. It means that the sensor data have to be delivered from WSN. One of the two proposed strategies can be applied to select the sensor nodes that have to be activated. Using the exhaustive strategy we activate all the eight sensor nodes ($i = 1,\ldots,8$) and we recognize the information granule that describe current positions of the prey with maximal available precision. In comparison, when using the minimum strategy we have to activate only two sensor nodes ($i = 4$ and $i = 5$) as they allow us to verify the possible state of the system (dark square), for which the optimal decision is to chase the second prey ($\hat{a}_t = 2$). For all the remaining states (light gray granules) the optimal control action is to chase the first prey $(\hat{a}_t = 1)$, thus according to the minimum strategy, these states do not have to be distinguished.

Tab. 1. Hypothetical states of the controlled system

| $\hat{I}(D)$ | $\hat{S}_t$ | $F(1,\hat{S}_t)$ | $F(2,\hat{S}_t)$ | $\hat{a}_t$ | $UNC(\hat{a}_t,\hat{S}_t)$ |
|---|---|---|---|---|---|
| (1,0,0,0,1,0,0,0) | (10,20)×(30,40) | (1,4) | (3,8) | 1 | 0.07 |
| (0,1,0,0,1,0,0,0) | (20,30)×(30,40) | (2,6) | (3,8) | 1 | 0.45 |
| (0,0,1,0,1,0,0,0) | (30,40)×(30,40) | (3,8) | (3,8) | 1 | 1.00 |
| (0,0,0,1,1,0,0,0) | (40,50)×(30,40) | (4,10) | (3,8) | 2 | 0.53 |
| (1,0,0,0,0,1,0,0) | (10,20)×(40,50) | (1,4) | (4,10) | 1 | 0.00 |
| (0,1,0,0,0,1,0,0) | (20,30)×(40,50) | (2,6) | (4,10) | 1 | 0.17 |
| (0,0,1,0,0,1,0,0) | (30,40)×(40,50) | (3,8) | (4,10) | 1 | 0.53 |
| (0,0,0,1,0,1,0,0) | (40,50)×(40,50) | (4,10) | (4,10) | 1 | 1.00 |
| (1,0,0,0,0,0,1,0) | (10,20)×(50,60) | (1,4) | (5,12) | 1 | 0.00 |
| (0,1,0,0,0,0,1,0) | (20,30)×(50,60) | (2,6) | (5,12) | 1 | 0.04 |
| (0,0,1,0,0,0,1,0) | (30,40)×(50,60) | (3,8) | (5,12) | 1 | 0.26 |
| (0,0,0,1,0,0,1,0) | (40,50)×(50,60) | (4,10) | (5,12) | 1 | 0.60 |
| (1,0,0,0,0,0,0,1) | (10,20)×(60,70) | (1,4) | (6,14) | 1 | 0.00 |
| (0,1,0,0,0,0,0,1) | (20,30)×(60,70) | (2,6) | (6,14) | 1 | 0.00 |
| (0,0,1,0,0,0,0,1) | (30,40)×(60,70) | (3,8) | (6,14) | 1 | 0.10 |
| (0,0,0,1,0,0,0,1) | (40,50)×(60,70) | (4,10) | (6,14) | 1 | 0.33 |

## 4. Information extraction for mobile object tracking

Object tracking is a popular application area of WSNs [22]. In this section an application is discussed, which uses the uncertainty based information extraction method for tracking of mobile object in WSN. The task of the considered object tracking application is to control the movement of a mobile sink which has to reach the target in the shortest possible time. On the basis of the proposed information extraction method three algorithms were developed that activates the sensor nodes and collects the sensor readings that are necessary to decide the movement direction of mobile sink.

Extensive simulation experiments were performed to compare performance of the proposed algorithm against results obtained for other methods that are available in the literature (i.e. prediction-based [23] and dynamic object tracking strategies [24]). The comparison was made with respect to information extraction cost and tracking performance. Hop counts and active times of sensor nodes were analyzed to evaluate the cost of information extraction from WSN. The control performance was measured as time-to-catch, i.e., the time in which the sink reaches the moving target.

It was assumed that the single target moves in a closed area, which is divided into square segments of equal dimensions. For each segment there is a sensor node available that can detect presence of the target. The control procedure is executed in discrete time steps. At each time step both the target and the sink move in one of the four directions: north, west, south or east. Their velocities (in segments per time step) are determined as parameters of the simulation. Target changes its movement direction randomly. Direction of the sink is decided by tracking algorithm on the basis of information extracted from WSN.

### 4.1. Tracking algorithms

In this study, five object tracking algorithms are compared. The first two algorithms were developed on the basis of the methods available in the literature. According to the first algorithm the target location is discovered at each time step and the sink is moved toward this location, i.e. such movement direction is selected which minimizes the distance between sink and target. The following action steps are executed for each time step, until target is reached:

Algorithm 1
1. Determine possible target locations.
2. Activate sensor nodes using exhaustive strategy.
3. Communicate target location to the sink.
4. Move sink toward target node.

Determination of the possible target locations is based on a simple model of the target movement, which is consistent with the above assumptions on available directions and predetermined velocity. Let us denote the target velocity by *vp*. If for previous time step ($t$–1) the target was detected in segment ($x$, $y$) then at time *t* there are four possible target locations: ($x$+ *vp*, $y$), ($x$–*vp*, $y$), ($x$, $y$+*vp*), and ($x$, $y$–*vp*). Application of the exhaustive strategy for sensor nodes selection means that the sensor nodes for all possible target locations have to be activated. Note that the target location is described by coordinates of a rectangular segment monitored by a single sensor node (target node).

The above algorithm utilizes the prediction-based object tracking method [23]. An important feature of this algorithm is that the extracted information about target location has the maximum available precision. Moreover, the information is delivered to the sink with the highest attainable frequency (at each time step of the tracking procedure).

The second examined algorithm was based on the tracking method which was proposed for the dynamical object tracking protocol (DOT) [24]. In this algorithm the location of target is discovered at each time step using the same approach as in Algorithm 1. The target location is then communicated to so-called beacon node. Each time the sink reaches the beacon node it gets information about current target location and the target node becomes new beacon node. Target node is the sensor node, which currently detects the target. At the beginning of the tracking procedure, the target location has to be discovered and communicated to the sink as the location of the first beacon node.

Algorithm 2
1. Determine possible target locations.
2. Activate sensor nodes using exhaustive strategy.
3. Communicate target location to the beacon node.
4. Move sink toward beacon node.
5. If beacon node is reached then beacon node := target node.

In case of algorithms 3 and 4, the proposed uncertainty-based information extraction method is employed in a straightforward manner. The difference between these algorithms lies in sensor nodes selection: algorithm 3 uses the exhaustive strategy, while algorithm 4 applies the minimal strategy to select the sensor nodes that have to be activated. Optimal movement direction for the sink is determined on the basis of predicted time-to-catch values.

The prediction of time-to-catch is made using the model of target movement. It takes into account possible locations and the known velocity of the target (*vp*) as well as of the sink (*v*). A potential future location of the sink $(x, y)_{dir}$ is determined for each movement direction (N, S, W, E), using the following formula:

$$(x,y)_{dir} = (x,y) + h(v_x, v_y)_{dir}, \tag{19}$$

where (*x*, *y*) denotes current position of the sink, *h* is a prediction horizon and *dir* indicates movement direction: $(v_x, v_y)_N = (0, v)$; $(v_x, v_y)_S = (0, -v)$; $(v_x, v_y)_E = (v, 0)$; $(v_x, v_y)_W = (-v, 0)$. For the proposed approach, the prediction horizon *h* is defined as a function of expected distance between sink and target (*dist*):

$$h = \lfloor \gamma \sqrt{dist} \rfloor, \qquad (20)$$

where $\gamma$ is a constant called prediction horizon parameter.

Distance between sink and target has to be described by an interval because the target location is determined as an information granule, i.e., a set of possible locations. The value of expected distance *dist* corresponds to the middle of this interval. Definition (20) allows us to obtain long prediction horizon in situations when the target is far from sink and shorter prediction horizons when the sink approaches the target. Note that *h* is determined in time steps of the control procedure.

As a result of the prediction, intervals are computed that describes time-to-catch for each of the four potential locations $(x, y)_{dir}$. On this basis, using the method described in Section 3, the optimal direction of movement is chosen.

The necessity of data transfer to the sink is recognized on the basis of expected decrease of uncertainty $\Delta UNC$. However, this rule does not take into account the current distance between sink and the possible target locations that are determined by means of prediction. Intuitively, in order to effectively catch the target, the actual target location should be checked when the sink is close to the predicted, potential target locations. Thus, additional heuristic rule in line 3 of Algorithm 3 was introduced to execute the data transfers whenever the sink is close to the area of possible target locations. The condition takes into account the variable *area*, which is defined as the number of possible target locations (segments). Due to the assumed target motion model, the area of possible target locations is a square shape. Therefore, the coefficient $dist/\sqrt{area}$ was used to describe relation between the expected distance to target and the side length of the rectangular area that includes possible target locations. Notice that this coefficient decreases if the sink gets closer to the possible target locations. The value $\beta$ is referred to as distance threshold.

Algorithm 3 (4)
1. Determine possible target locations.
2. For all movement directions predict time-to-catch.
3. If $\Delta UNC > \alpha$ or $dist/\sqrt{area} < \beta$ then
3.1. activate sensor nodes using exhaustive (minimum) strategy.
3.2. If target is detected communicate its location to the sink.
4. Determine optimal movement direction.
5. Move sink in the optimal direction.

Algorithm 5 combines the proposed uncertainty-based method with the approach of the DOT protocol. It consists of two parts that are executed concurrently – one at the target node and second at the sink. Sink determines possible locations of the target and predicts time-to-catch for each movement direction. If at a given time step such prediction is insufficient to provide acceptable level of the decision uncertainty or the sink is close to the possible target locations then the sink sends request to the beacon node and gets the actual target location. At each time step of the control procedure the information about current location of the target is delivered to the beacon node.

Algorithm 5
At target node:

1.  Determine possible target locations.
2.  Activate sensor nodes using exhaustive strategy.
3.  Communicate target location to the beacon node.

At sink:
1.  Determine possible target locations
2.  For all movement directions predict time-to-catch
3.  If $\Delta UNC > \alpha$ or $dist/\sqrt{area} < \beta$ then
3.1.    get location of target node from beacon node,
3.2.    beacon node := target node.
4.  Determine optimal movement direction.
5.  Move sink in the optimal direction.

### 4.2. Experimental results

The above object tracking algorithms for WSNs were evaluated in simulation experiments. Results of the simulations include hop counts, total active times of sensor nodes, number of data transfers to sink and time-to-catch values. Both the active time and the time-to-catch are measured in time steps of the control procedure. Hop counts were determined assuming that the shortest path is used for each data transfer. All sensor nodes have the same communication range covering the nearest four nodes in the neighboring segments. The monitored area was assumed to be a square of 200 x 200 segments. Therefore, the number of sensor nodes equals 40 000.

Due to the large size of the network, the use of standard simulation platforms (NS2 and OPNET) has caused excessive simulation time and memory consumption. Thus, the experiments were performed using simulation software that was developed for this research. Accuracy of the simulation software was verified by comparison to results obtained from NS2 and manual calculations for a comprehensive set of scenarios. The simulations in NS2 were performed for smaller number of sensor nodes (up to 5000).

Initial experiments were carried out in order to examine influence of the parameters $\alpha$, $\beta$ and $\gamma$ on performance of Algorithms 3-5. The results in Figs. 2 and 3 illustrate the effect of these parameters on transmission cost as well as on tracking performance. These results were obtained using Algorithm 5, for target velocity equal 3 and sink velocity equal 4. In this study, all velocities are expressed in segments per time step. The results were averaged for 100 simulation runs. Each simulation run starts with the same locations of sink and target. During simulation, random trajectory of the target is generated. The simulation stops when target is caught by the sink.

Charts in Fig. 2 illustrate the influence of parameters $\alpha$ (the threshold of expected uncertainty decrease) and $\beta$ (the prediction horizon parameter). Black color in these charts corresponds with low level of the analyzed quantities. The upper left plot shows that high hop count values are obtained for $\alpha$ below 0.1 as well as for $\beta$ above 4. This effect is caused by the more frequent data transfers to sink. The high hop counts for $\alpha$ above 0.15 and $\beta$ below 0.5 are related to the long time-to-catch that was obtained for such settings (see the upper right plot). Based on the results presented in Fig. 2, the optimal values of parameters were determined: $\alpha = 0.15$ and $\beta = 2.0$. These settings were used for all the simulations reported in this section.

Impact of the parameter $\gamma$ (the prediction horizon parameter) is analyzed in Fig. 3. The best performance of the tracking algorithm corresponds with the lowest time-to-catch, which was observed for $\gamma = 1.3$. Minimal hop count and total active time of sensor nodes were also registered for $\gamma = 1.3$. Therefore, this setting was chosen as the optimum for further experiments.

The results discussed above concern the performance of Algorithm 5, however for Algorithms 3 and 4 similar findings were obtained. Thus, the same values of the analyzed parameters were applied in our experiments for Algorithms 3, 4, and 5.

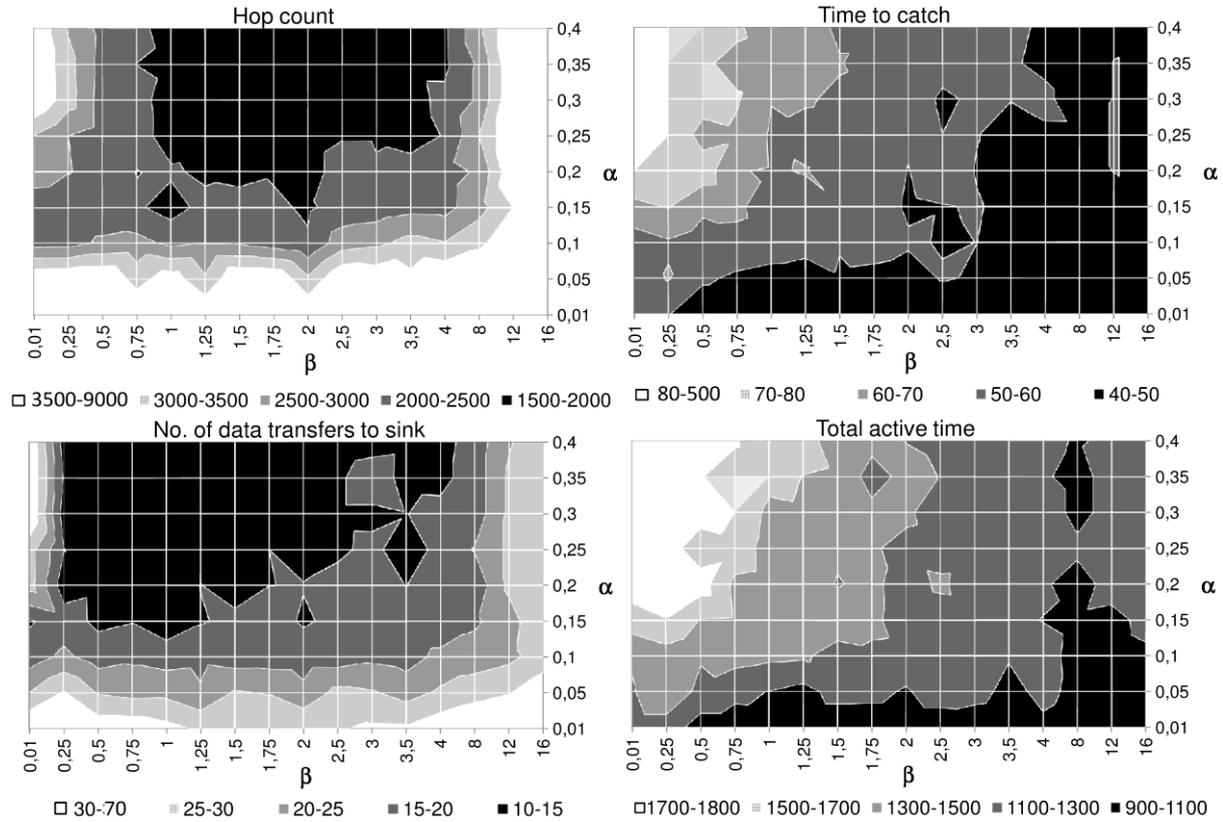

Fig. 2. Transmission cost and tracking performance for different values of threshold of expected uncertainty decrease $\alpha$ and distance threshold $\beta$

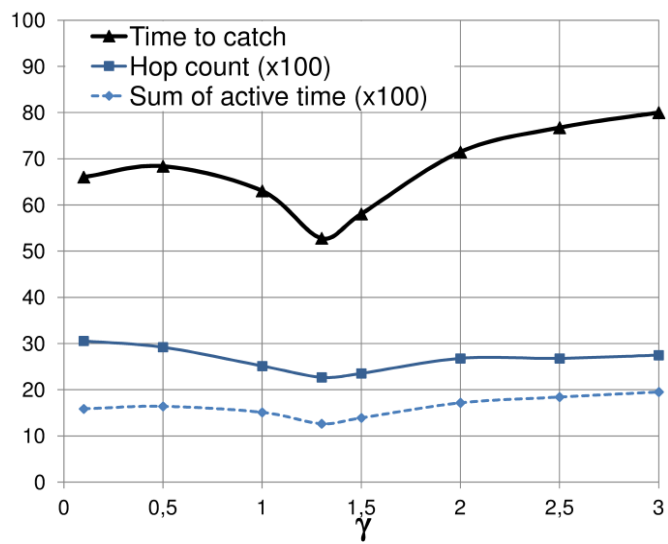

Fig. 3. Impact of prediction horizon parameter $\gamma$ on tracking performance and communication cost.

In the sequel, comparison is made between the five examined algorithms taking into account their performances for three different velocities of the target ($vp = 1\ldots3$). Velocity of the sink was equal 4 in each simulation. One point that is immediately obvious from the results presented in Figs. 4 and 5 is that the three proposed algorithms (Alg. 3 – 5) require low number of data transfers and provide short time-to-catch values. In comparison, Algorithm 1 involves significantly higher number of data transfers and Algorithm 2 needs longer time to reach the moving target than the other considered algorithms. For Algorithms 1, 3, 4, and 5 the difference of time-to-catch values is lower than 10 time steps. As it could be expected, the shortest time-to-catch was obtained for Algorithm 1, in which the extracted information about target locations has the highest available precision.

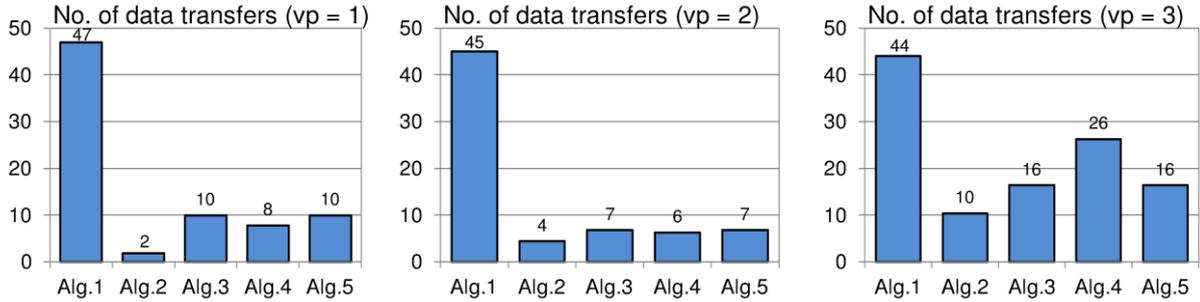

Fig. 4. Number of data transfers to sink for different target velocities

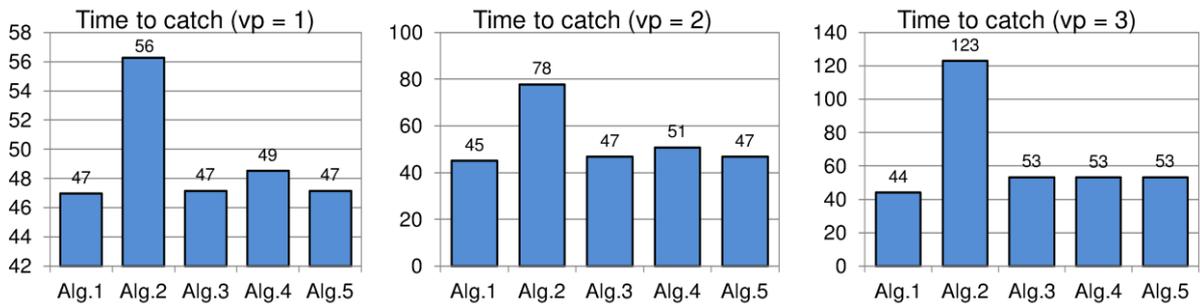

Fig. 5. Time-to-catch (in time steps) for different target velocities

Fig. 6. shows hop counts that were registered during the experiments. Hop count is a highly important factor for WSNs applications because it strongly influences the amount of energy consumed by sensor nodes. Total active times of sensor nodes are presented in Fig. 7. During active time the energy is consumed by the sensing subsystem of a sensor node. It depends on the type of applied sensors if the sensing subsystem is a significant source of energy consumption [25]. When analyzing the data illustrated in Figs. 6 and 7 it can be observed that for most cases the lowest data transmission cost was obtained by Algorithm 5, which is based on the method introduced in Section 3.

The two remaining algorithms that apply the proposed method (Algorithm 3 and Algorithm 4) result in highest transmission costs. These results are due to the fact that better solution for target tracking is to recognize the target location at each time step than to skip the target detection when this information is not requested. In the analyzed system, when target is detected at each time step over a period of $T$ time steps, sensor nodes have to be activated $4T$ times (assuming that $vp = 1$ and the initial target location is known). However, if we try to find the target location once after $T$ time steps then we need to activate $0.5[(2T+1)^2-1]$ sensors. E.g., for $T = 5$ the number of sensor activations equals 20 in the first case and 60 in the second case. This difference rises rapidly for higher target velocities. Therefore, the reduction in frequency of data transfers for Algorithm 3 and Algorithm 4 is obtained at the expense of higher utilization of sensor nodes. Above insights have motivated the solution used in Algorithm 5, in which the target location is recognized at each time step but the sink gets this information only if it is necessary for decision making.

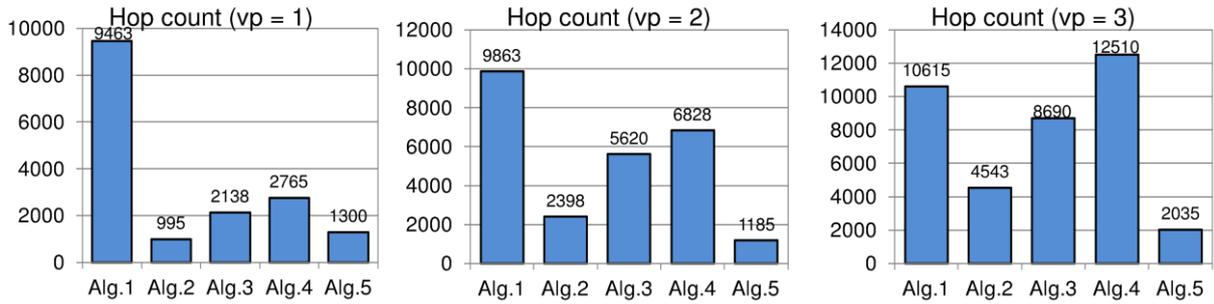

Fig. 6. Hop count for different target velocities

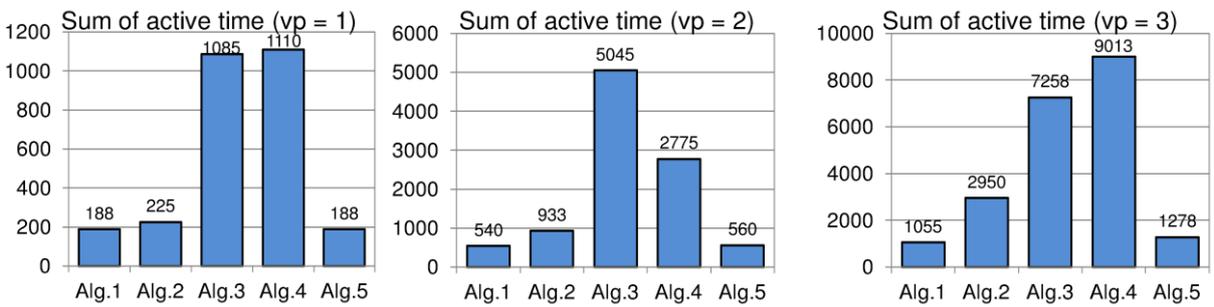

Fig. 7. Total active time of sensor nodes (in time steps) for different target velocities

According to the presented results, it could be concluded that Algorithm 5, which is based on the proposed information extraction method, enables a significant reduction of the data transmission cost and ensures good performance of the tracking application. The comparison of simulation results for Algorithms 1 and 5 shows that Algorithm 5 reduces hop count by about 85% and increases time-to-catch by 8% on average.

An example of a simulated target trajectory and resulting sink trajectories are presented in Fig. 8. The *x*- and *y*-axis correspond to the discrete coordinates (*x*, *y*) of the segments monitored by sensor nodes. The different trajectories of sink were obtained by using Algorithm 5 and the algorithms taken from literature (Algorithm 1 and Algorithm 2). Velocity of the target was 2 segments per time step and velocity of the sink was 4 segments per time step in this example. At the beginning of the simulation, target is located in segment with coordinates (66, 66) and sink is in segment (160, 160). For Algorithm 1 the target is caught after 32 time steps in segment (128, 66). When using Algorithm 5 the sink reaches target in 37-th time step at segment (136,66). Algorithm 2 requires 89 time steps and finds the target in segment (188, 116).

Fig. 9 shows the course of changes in both hop count and total active time of sensor nodes for the analyzed tracking example. These results confirm advantages of the introduced algorithm. The left plot in Fig. 9 shows that the fastest growth of hop count was observed for Algorithm 1 at the beginning of the tracking procedure and it was only slightly decreased when the sink was approaching the target. It means that Algorithm 1 generate relatively high data traffic in WSN during entire period of tracking. In comparison, for Algorithm 5 the hop count grows quite slowly at the beginning and faster when the target gets closer. At time steps 1-22 growth of the hop count for Algorithm 5 is similar to that obtained for Algorithm 2, however the total hop count for Algorithm 2 is significantly higher (close to the result of Algorithm 1) due to the long time-to-catch.

For the three algorithms that are compared in Fig. 9 the same number of sensor nodes has to be activated at each time step to detect the target until the target is caught. As the lowest time to catch is obtained for Algorithm 1, the total active time is also lowest for this algorithm. Algorithm 1 needs the largest number of hops because the information about target position is transmitted to the sink at each time step of the tracking procedure. For Algorithms 2 and 5 the hop count is lower because at each time step the target position is reported to the beacon node, which is closer to the target than the sink. According to the results presented in Fig. 9, Algorithm 5 provides the lowest data transmission cost as it needs significantly lower number of hops and its total active time is very close to the result of Algorithm 1.

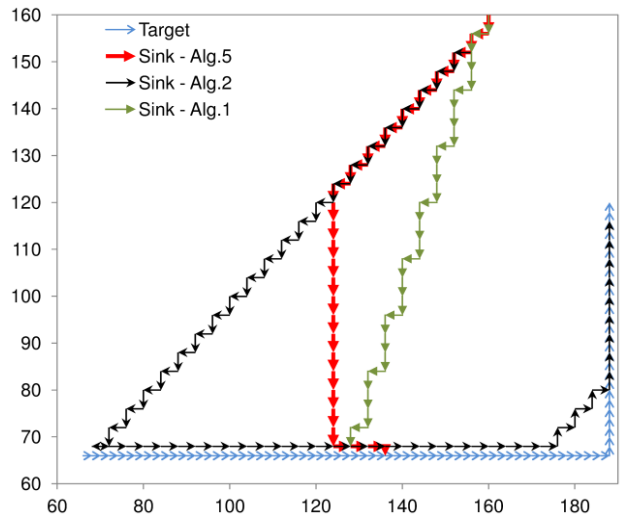

Fig. 8. Target trajectory and sink trajectories for compared tracking algorithms

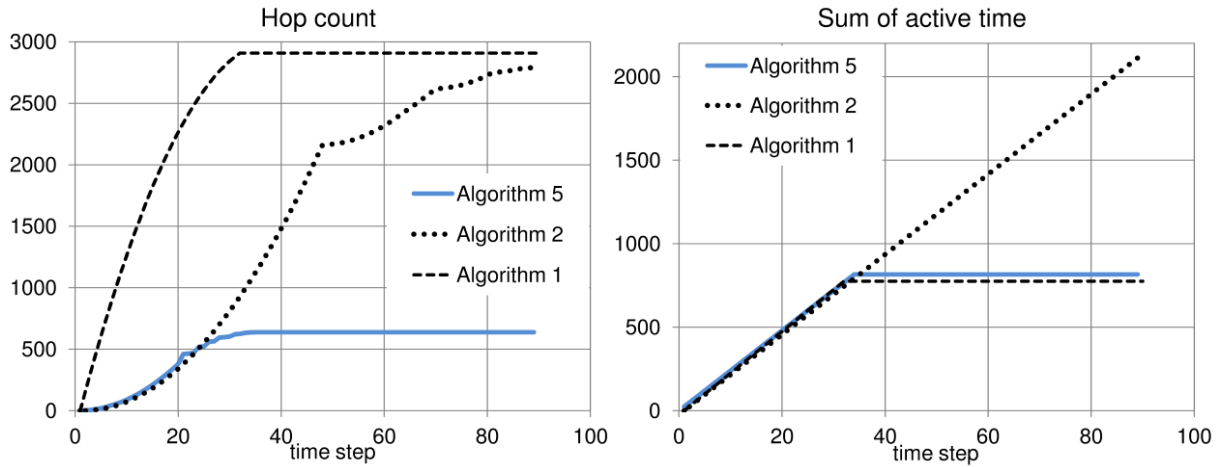

Fig. 9. Hop count and total active time of sensor nodes for compared tracking algorithms

It should be noted here that the target trajectory is generated randomly during simulation. Fig. 10 gives a sample of the simulation results for complex target trajectories. It illustrates the diversity of the randomly generated target trajectories that were taken into account during experiments. As it can be observed in these examples, the generated trajectories have included reverse paths and loops. Reverse path is the case when target turns back (upper right part of the right chart in Fig. 10). Loop is shown in lower left corner of the left chart. It can be also observed in these charts that tracks of the sink for Algorithms 1 and 5 are considerably shorter than for Algorithm 2. Algorithm 5 uses different paths than Algorithm 1, however the time-to-catch for these two algorithms is comparable.

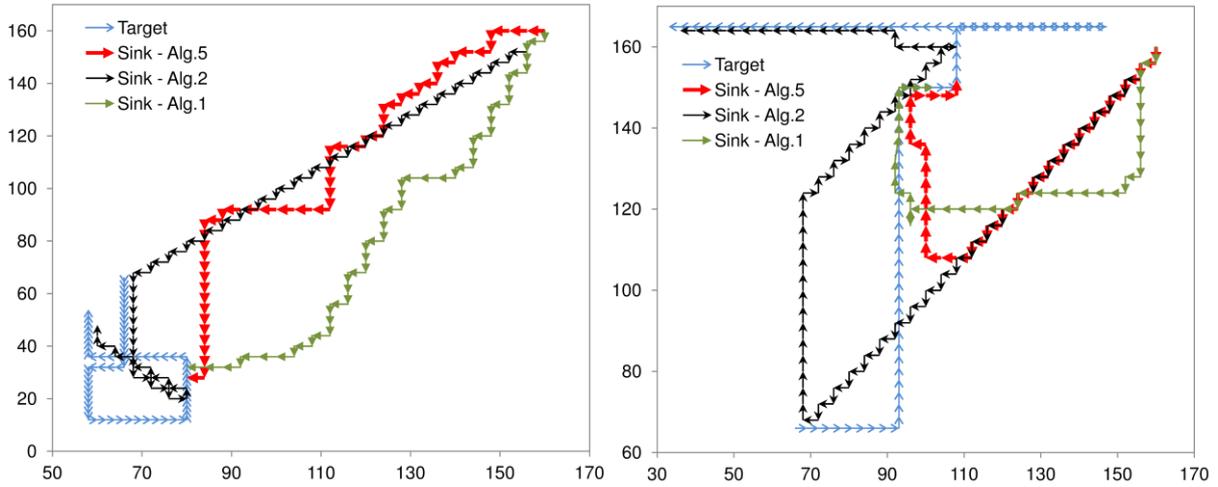

Fig. 10. Examples of object tracking (target and sink trajectories)

## 5. Conclusions and future work

Specific requirements of the WSNs control applications necessitate new methods for information extraction. In order to optimize the utilization of WSN, the scope and precision of the extracted information have to be dynamically adjusted to the variable needs of control decisions. The method proposed in this paper analyzes uncertainty of control decisions to extract information which is necessary and sufficient for achieving control objectives. Using this method a control application can both recognize situations when new data have to be collected and determine sensor nodes that have to be activated. The concept of information granules was applied in this approach to provide a flexible state description of the controlled system. Precision of the granular description is changeable and can be easily adapted to the set of delivered input data.

The proposed uncertainty-based information extraction method is general in many aspects and can therefore be used for a large class of control applications. First of all, the method does not assume any specific type of the model of controlled system or prediction technique. It can be implemented using different formal frameworks for the representation of information granules. Moreover, it can be easily combined with other information extraction approaches that are available in literature, e.g. data aggregation and data suppression.

In this paper the proposed method was used to develop new algorithms for object tracking in WSNs. The task under consideration is to control movement of a mobile sink, which has to reach a target in the shortest possible time. Performance of the proposed approach was compared against state-of-the-art methods. The experimental results show that the introduced information extraction method enables substantial reduction in the amount of transmitted data with no significant negative effect on tracking performance. The most promising results were obtained for combination of the proposed method with the prediction-based in-network tracking. It was also observed that the results can be enhanced by extending the uncertainty-based method with heuristic rules that takes into account specific character of the particular control task.

Results of this study allow us to identify several areas for further research. One of the intriguing problems is the sensor nodes selection, which should take into account the uncertainty and cost of data transmission predicted for a time horizon longer than one time step of the control procedure. Another issue is related to decomposition of the uncertainty evaluation, which will enable distributed execution of the necessary computations at multiple sensor nodes. Note that according to the algorithms discussed in this study, the vast majority of calculations have to be conducted by the sink. An interesting topic for further research is also the comparison of control performance and data transmission cost for applications based on different formal representations of the information granules.